\documentclass[vecphys]{svmult}

\usepackage{makeidx}         % allows index generation
\usepackage{graphicx}        % standard LaTeX graphics tool
\usepackage{multicol}        % used for the two-column index
\usepackage[bottom]{footmisc}% places footnotes at page bottom
\makeindex             % used for the subject index
\begin{document}

\title*{Evolution of the Small Magellanic Cloud}
\author{Kenji Bekki\inst{1}}
\institute{School of physics,
University of New South Wales,
Sydney 2052,
Australia
\texttt{bekki@phys.unsw.edu.au}}
\maketitle

%Your text goes here. Separate text sections with the standard \LaTeX\
%sectioning commands.

\section{Numerical simulations of the SMC evolution}
\label{sec:1}
% Always give a unique label
% and use \ref{<label>} for cross-references
% and \cite{<label>} for bibliographic references
% use \sectionmark{}
% to alter or adjust the section heading in the running head
%%%Your text goes here. Use the \LaTeX\ automatism for your citations

We investigate (1) the origin of the bifurcation of the Magellanic
stream (MS), (2) the formation of distinctively metal-poor
stellar populations in the Large Magellanic Cloud (LMC)
 due to sporadic gas transfer from
the SMC, and (3) the collision between the leading arms (LAs) of
the MS and the outer part of the Galactic HI disk based on
numerical simulations of LMC-SMC-Galaxy interaction for the last
2.5 Gyr (e.g., Bekki \& Chiba 2007, BC07).
We adopt  numerical methods and techniques of the simulations
on the evolution of the MCs
used in our previous papers (Bekki \& Chiba 2005; B05):
we first determine the most
plausible and realistic orbits of the MCs 
by using `` the backward integration
scheme'' (for orbital evolution  of the MCs) by Murai \&  Fujimoto (1980)
for the last 2.5 Gyr and then investigate the
evolution of the MCs using GRAPE
systems (Sugimoto et al.1990).
The total masses of the LMC  and the SMC 
are set to be $2.0 \times 10^{10} {\rm M}_{\odot}$
and $3.0 \times 10^{9} {\rm M}_{\odot}$,
respectively,  in all models.
The SMC is represented by a fully self-consistent dynamical
model wheres
the LMC is represented by a point mass.
We adopt the initial locations and velocities
of the MCs  with respect to
the Galaxy that are similar to those adopted by BC07.

%\cite{monograph}.

% Use the \index{} command to code your index words

\begin{figure}
\centering
\includegraphics[height=8cm]{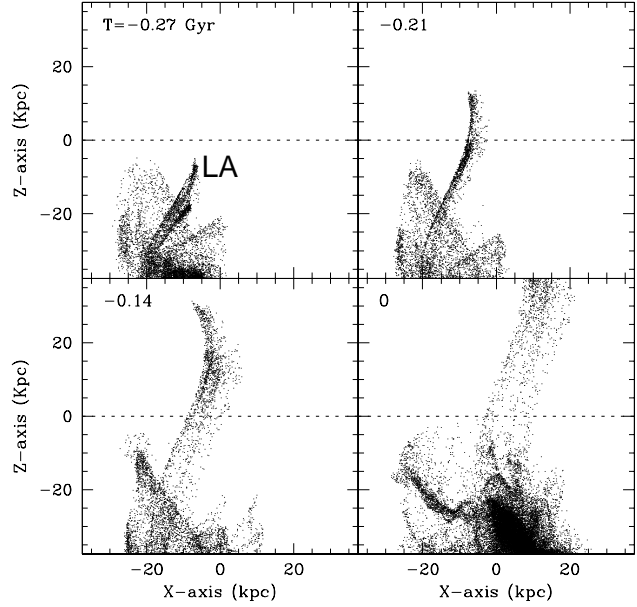}
\caption{
Time evolution of the LAs of the MS  for the last 0.27 Gyr.
Only particles within 70 kpc from the center of the Galaxy
are shown so that only the particles in the LAs can be 
more clearly seen. The time $T=0$ Gyr  means the present whereas
$T=-0.27$ Gyr means 0.27 Gyr ago. The dotted line represents
the disk plane of the Galaxy. Note that the LAs are composed of
two streams passing through the Galactic disk about 0.2 Gyr ago.
}
\label{fig:1}       % Give a unique label
\end{figure}

\begin{figure}
\centering
\includegraphics[height=6cm]{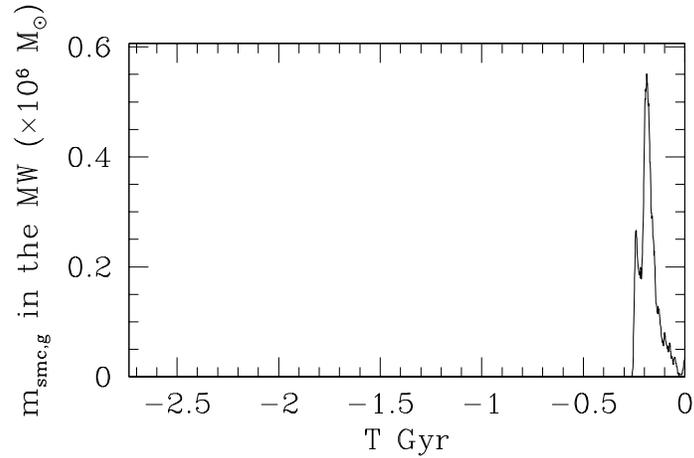}
\caption{
Time evolution of the total mass of the SMC's gas particles
that are within the central 35 kpc of the Galaxy
($m_{\rm SMC,g}$). The locations of the peaks
correspond to the epochs when  the LAs pass through
the Galactic disk.}
\label{fig:2}       % Give a unique label
\end{figure}

\begin{figure}
\centering
\includegraphics[height=6cm]{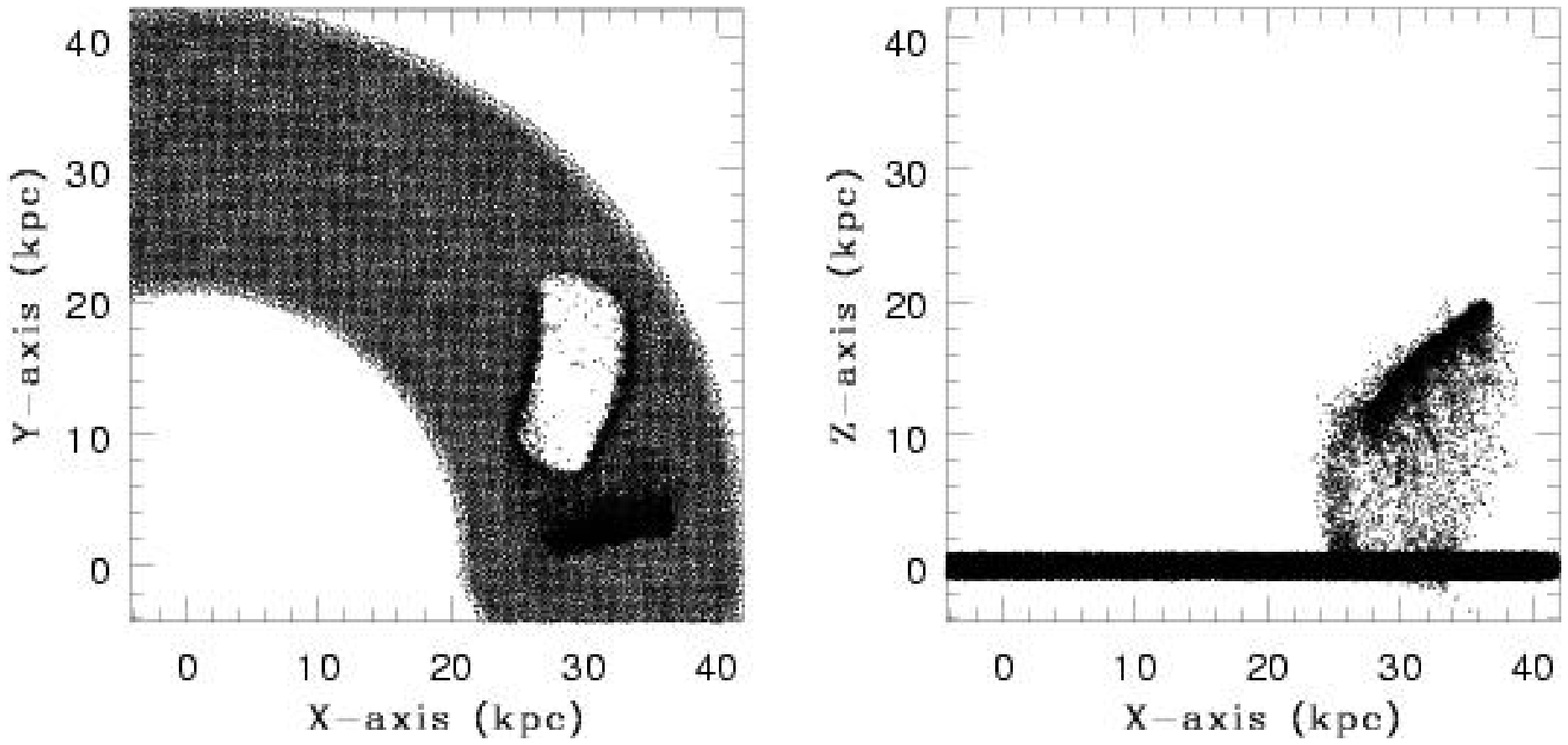}
\caption{
The final gaseous distributions projected onto the $x$-$y$ plane
(left) and the $x$-$z$ one for the Galactic HI disk about 0.2 Gyr
after the Magellanic impact. Note that as a result of the Magellanic
impact, a giant HI hole can be created in the Galactic HI disk.}
\label{fig:3}       % Give a unique label
\end{figure}

\section{The Magellanic impact}
\label{sec:2}

Although the bifurcation of the MS (i.e., two streams running parallel
with each other in the MS) can be reproduced by the present 
tidal interaction model of the MS formation
as well as by the previous ones (e.g., Connors et al. 2006),
it has been unclear whether the tidal models can also reproduce 
the location of the kink of the LAs in the MS. We here propose that
the observed location of the kink can be reproduced by the tidal model
if the hydrodynamical interaction between the LAs and the outer part of
the HI disk of the Galaxy is modeled in a reasonable and realistic way.
Fig. 1 clearly shows
that the LAs, which are composed of two main streams, can pass through
the outer part of the Galactic HI disk about 0.2 Gyr.  
This collision between the LAs and the HI disk is referred to as
``the Magellanic impact'' from now on for convenience.
Fig. 2 shows that there are two peaks in the time evolution
of the total mass ($m_{\rm SMC,g}$)
of the SMC's gas particles that were initially
within the SMC and later stripped to be {\it temporarily} within
the central 35 kpc of the Galaxy (note that particles 
within the central 35 kpc are counted or not at each time step,
regardless of whether they are already counted prior to the time  step).
About 1\% of the initial gas mass of the SMC 
(which corresponds to an order of $10^7 {\rm M}_{\odot}$) can pass through
the HI disk of the Galaxy within the last 0.2 Gyr during
the Magellanic impact.

Fig. 3 shows the results of our hydrodynamical simulations
on the collision between the LA and the outer part of the HI disk
of the Galaxy. The final snapshot of the simulation shown in Fig. 3
demonstrates that the Magellanic impact can push out some fraction
of the HI gas of the Galaxy so that a giant (kpc-scale) HI hole
can be created about 0.2 Gyr after the Magellanic impact.
A chimney-like bridge connecting between the LA and the Galactic HI
disk can be also created by the Magellanic impact. The HI velocity
field close to the giant hole is significantly disturbed so that
ongoing and future observations on kinematics of the Galactic HI 
disk can detect such a disturbance. Furthermore, the high-density
ridge along the giant hole is  one of the predicted properties that can
be tested against  observations. Owing to the hydrodynamical interaction
between the Galactic HI disk and the LA, the orbit of the LA can
be significantly changed after the Magellanic impact at $l=0$.
This means that the location of the simulated kink of the LA
should be around $l=0$, which is consistent with
the observed location of the kink.

\section{The Magellanic squall} 
\label{sec:3}

One of another important results of the present simulations is 
that a significant  amount of metal-poor gas is stripped from
the SMC and fallen into the LMC
during the LMC-SMC-Galaxy  interaction over the last 2 Gyrs. 
We find  that the LMC can temporarily replenish gas supplies
through the sporadic accretion and infall   of metal-poor gas from the SMC.
We also find that about 0.7 \% and 18 \%  of the SMC's gas
can pass through the central region of the LMC about 1.3 Gyr ago
and 0.2 Gyr ago, respectively.
The possible mean metallicity of the replenished gas from the
SMC to LMC is about [Fe/H] = -0.9 to -1.0 for the two
interacting phases for an adopted steep initial metallicity
gradient of the SMC's gas disk (BC07).
We thus propose that this metal-poor gas
can closely be associated with the origin of the observed  LMC's young and
intermediate-age stars and star clusters with distinctively
low-metallicities with [Fe/H] $< -0.6$
(e.g., Grocholski et al. 2006; Santos \& Piatti 2004).
We also suggest that if these gas from the SMC can collide with
gas in the LMC to form new stars in the LMC, the metallicities of
the stars can be significantly lower than those
of  stars formed from
gas initially within the LMC.
Accordingly this ``Magellanic squall'', which means gas-transfer from the SMC
to the LMC, can be very important for the recent star formation
history of the LMC: The evolution of the LMC is influenced not only
by tidal effects of the SMC and the Galaxy but also by the mass-transfer
from the SMC, if the MCs  have been interacting with each other
for the last 2 Gyrs.

%
% For built-in environments use
%
%\input{referenc}

%%%%%%%%%%%%%%%%%%%%%%%%%%%%%%%%%%%%%%%%%%%%%%%%%%%%%%%%%%%%%%%%%%%%%%  }

%%%%%%%%%%%%%%%%%%%%%%%%%%%%%%%%%%%%%%%%%%%%%%%%%%%%%%%%%%%%%%%%%%%%%%

\printindex
\end{document}